\documentclass[aip,reprint]{revtex4-1}

\draft 
\usepackage{graphicx}

\begin{document}


\title{Reduced Dark Counts in Optimized Geometries for Superconducting Nanowire Single photon Detectors} 



\author{Mohsen K. Akhlaghi}
\email[]{mkeshava@maxwell.uwaterloo.ca}
\affiliation{ECE Department, University of Waterloo, 200 University
Ave West, Waterloo, ON, Canada, N2L 3G1}

\author{Haig Atikian}
\affiliation{School of Engineering and Applied Sciences, Harvard
University, Cambridge, MA 02138}

\author{Amin Eftekharian}
\affiliation{ECE Department, University of Waterloo, 200 University
Ave West, Waterloo, ON, Canada, N2L 3G1} \affiliation{Institute for
Quantum Computing, University of Waterloo, 200 University Ave West,
Waterloo, ON, Canada, N2L 3G1}

\author{Marko Loncar}
\affiliation{School of Engineering and Applied Sciences, Harvard
University, Cambridge, MA 02138}

\author{A. Hamed Majedi}
\email[]{ahmajedi@maxwell.uwaterloo.ca} \affiliation{ECE Department,
University of Waterloo, 200 University Ave West, Waterloo, ON,
Canada, N2L 3G1} \affiliation{Institute for Quantum Computing,
University of Waterloo, 200 University Ave West, Waterloo, ON,
Canada, N2L 3G1} \affiliation{School of Engineering and Applied
Sciences, Harvard University, Cambridge, MA 02138}


\date{\today}

\begin{abstract}
We have experimentally compared the critical current, dark count
rate and photo-response of 100nm wide superconducting nanowires with
different bend designs. Enhanced critical current for nanowires with
optimally rounded bends, and thus with no current crowding, are
observed. Furthermore, we find that the optimally designed bend
significantly reduces the dark counts without compromising the
photo-response of the device. The results can lead to major
improvements in superconducting nanowire single photon detectors.
\end{abstract}


\maketitle 


Single photon detectors are essential components in diverse fields
including quantum optics and information \cite{Knill2001A}, quantum
key distribution \cite{Takesue2007Qu}, lunar laser communication
\cite{Grein2011De}, diagnosis of integrated circuits
\cite{Zhang2003No} and characterization of single photon sources
\cite{Stevens2006Fa}. Superconducting nanowire single photon
detectors (SNSPDs) outperform other detectors in merits such as
infrared quantum efficiency, dark count rate, timing jitter
\cite{Hadfield2009S}, and maximum count rate \cite{Akhlaghi2012Ga}.
Thus, they are considered as a promising technology for demanding
photon counting applications \cite{Natarajan2012Su}.

SNSPDs are typically made of current biased meandering
superconducting nanostrips (usually $\sim$100nm wide) with
180-degree turns. The photons are focused on the parallel nanostrips
that form the active area, while the turns only serve the purpose of
electrical connection. The closer the bias current is to the
critical current of the nanostrips, the higher the detection
efficiency, but also the higher the dark count rate
\cite{Natarajan2012Su}. Although the turns are typically placed
outside the photon absorbing area, and thus do not directly
contribute to the photon detection, they can degrade the overall
performance of the detector by acting as current bottlenecks or by
generating dark counts.

Recently, Clem et al. \cite{Clem2011Ge} recaped the possible impact
of sharp turns on SNSPDs: the current crowds at the inner edge thus
reducing the measured critical current of the meander. Also, the
current bottleneck in wide superconducting strips (300nm to 1$\mu$m
wide) with sharp bends has been experimentally demonstrated
\cite{Hortensius2012Cr, Henrich2012Ge}. However, an open question
remains on the impact of current crowding on present SNSPDs that
feature much narrower strips ($\sim$100nm wide), in which both
increased ratio of the bend curvature (due to inherent finite
fabrication resolution) to nanowire width, and reduced width to
coherence length ratio make the expected effect smaller
\cite{Clem2011Ge, Henrich2012Ge}. Here we present experiments that
probe the current crowding effect on the critical current of
superconducting nanostrips with a width comparable to the commonly
used width in modern SNSPDs. We also report on the effect of sharp
bends on the observed photo-response and dark counts.

\begin{figure}[t]
\centering\includegraphics[width=8.0cm]{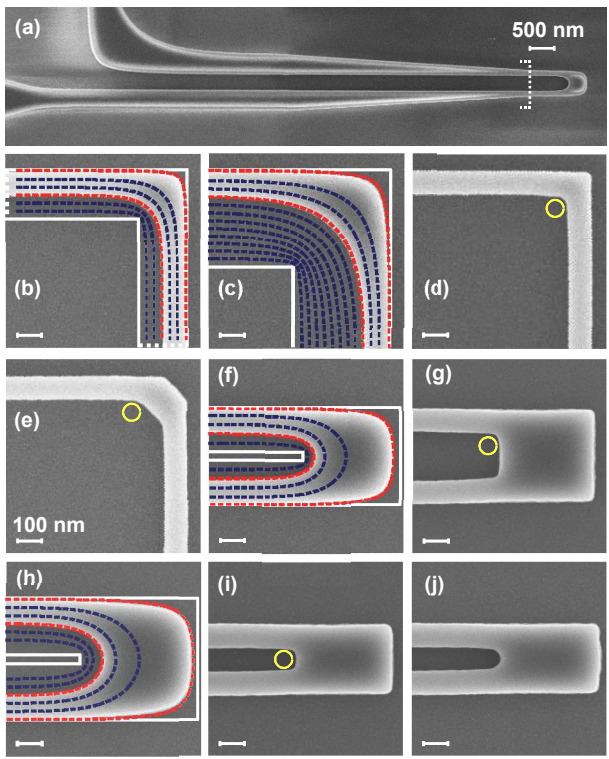} \caption{\label{fig:1}
(Color online) Scanning electron microscope images of the nanowires
explored in this letter. (a) A typical nanowire structure examined
in this letter and its connection lines. (b) and (c) two optimized
90-degree bends. (d) and (e) sharp and 45$^{\circ} $ 90-degree
bends. (f) and (g) optimized and sharp 180-degree turns with 200nm
spacing. (h) optimized 180-degree turn with 300nm spacing. (i) and
(j) sharp and circular (radius = 50nm) 180-degree turns with 100nm
spacing. The circles are eye guides with 35nm radius. Blue and red
dashed lines are current streamlines calculated for a superconductor
thin film enclosed by solid white lines. All the parts, except (a)
share the same length scale.}
\end{figure}

A typical device presented in this letter is illustrated in
Fig.~\ref{fig:1}(a). A nanowire, 100nm wide and 8nm thick, is bent
either 90-degree or 180-degree, and connected to large pads (not
shown) by a gradual transition to wider strips. The nanowire length
is $\sim$0.5$\mu$m on either side of the bend. Our bends fall into
two categories: optimally designed with no current crowding and thus
no expected critical current reduction, and traditional bends made
without optimal considerations.

Figure~\ref{fig:1}(b) shows an example of our optimum bends.
\nolinebreak{To find the optimal bend design, we numerically solve
$\nabla \cdot \textbf{\textit{K}}=0$ and $\nabla \times
\textbf{\textit{K}} = -d/\lambda^2 \textbf{\textit{H}} \approx0$}
within the area enclosed by the white lines \cite{Clem2011Ge}, where
$\textbf{\textit{K}}$ is the sheet current density, $d$ is the
nanowire thickness and $\lambda$ is the magnetic penetration depth.
The boundary conditions are $\textbf{\textit{n}} \cdot
\textbf{\textit{K}}_{l} = 0$, and $\textbf{\textit{n}} \times
\textbf{\textit{K}}_{i} = 0$, where $\textbf{\textit{n}}$ is a
vector normal to the edge, $\textbf{\textit{K}}_{l}$ is
$\textbf{\textit{K}}$ on the lateral boundaries (solid white lines)
and $\textbf{\textit{K}}_{i}$ is $\textbf{\textit{K}}$ on the input
boundaries (dotted white lines). Next, we find the streamlines of
the vector field $\textbf{\textit{K}}$ (dashed blue and red lines).
Any two streamlines (dashed red lines) that enclose a surface within
which $|\textbf{\textit{K}}|$ remains less than or equal to
$|\textbf{\textit{K}}_{i}|$, form an optimized bend (because
$\textbf{\textit{K}}$ within them satisfies the same above boundary
value problem, and $|\textbf{\textit{K}}|$ in the bend does not
exceed $|\textbf{\textit{K}}|$ within the nanowire).

Four different 90-degree bends have been investigated: (i) optimized
bend with the smallest possible footprint, (ii) optimized bend twice
as big as the smallest one (to make it more tolerant to fabrication
errors), (iii) sharp bend and (iv) 45$^{\circ}$ bend (as a structure
between worse and best case scenarios) (see Figs.~\ref{fig:1}(b)
through ~\ref{fig:1}(e)). The smallest possible optimum 180-degree
turn (200nm spacing) is shown in Fig.~\ref{fig:1}(f). It will be
compared with a sharp 180-degree turn (200nm spacing) and a bigger
optimum turn (300nm spacing) as shown in Figs.~\ref{fig:1}(g) and
~\ref{fig:1}(h). Finally, Figs.~\ref{fig:1}(i) and ~\ref{fig:1}(j)
present a commonly used bend in present SNSPDs (sharp bend with
100nm spacing) and the same but circularly rounded (radius = 50nm).

The devices are made of 8nm thick NbTiN films deposited on oxidized
silicon. Hydrogen silsesquioxane resist was spin-coated on top and
pattered using 125keV electron-beam lithography. The write
parameters were carefully tuned to achieve nanostructures as
identical as possible to the designed curvatures (see red dashed
lines in Fig.~\ref{fig:1} overlayed on the nanowire images). The
resist was developed in a tetra-methyl ammonium hydroxide solution,
and the pattern was transferred into the film using ion beam milling
with Argon gas. The critical temperature of the film before and
after nano-patterning was measured to be ~8.4K

The critical current, dark count and photo response of a nanowire is
a function of its dimensions (thickness and width), as well as
superconducting thin film quality. Therefore, when investigating the
effect of bend design, it is essential to keep the nanowires
identical except at the bend. In our experiments, we only juxtapose
two different bends from the designs in Fig.~\ref{fig:1} that
satisfy the following conditions: (i) both are either 90-degree or
180-degree, and (ii) both are fabricated few $\mu$m apart on the
same chip. The first condition keeps the two geometries as similar
as possible and therefore minimizes slight width changes when
different geometries are exposed by the electron-beam. The second
condition assures the two nanowires share the most identical film
thickness/quality as well as equivalent fabrication processing (to
make effects of many factors including resist variations, proximity
dose effects, and others less significant).

Each pair of nanowires has a common electrical ground. Each of the
other terminals connect to a 490nH inductor (placed next to the
chip) and then to a room temperature bias-T by a coax cable
(50$\Omega$ impedance). A computer controlled voltage source that
measures its output current (Keithley 2400) is connected to the DC
port of the bias-T via a low-pass filter (to reduce high frequency
noise and interference). The high frequency response of the nanowire
(after room temperature amplification) is monitored through the RF
port on an oscilloscope or a programmable counter. A single mode
fiber, placed several centimeters away from the chip uniformly
radiates the pair with 1310nm photons from an attenuated pulsed
laser source (width $\sim$200ps, repetition rate 20MHz). The
50$\Omega$ impedance together with the inductor make a large enough
time constant to observe relaxation oscillations in all our
current-voltage curve measurements, thus ensuring the peak current
is the (experimental) critical current \cite{Kerman2009El}. The
measurements have been done by installing the samples in a dipstick
probe and immersing it in liquid Helium (monitored temperature
$\sim$4.2K).

\begin{figure}[t!]
\centering\includegraphics[width=8.0cm]{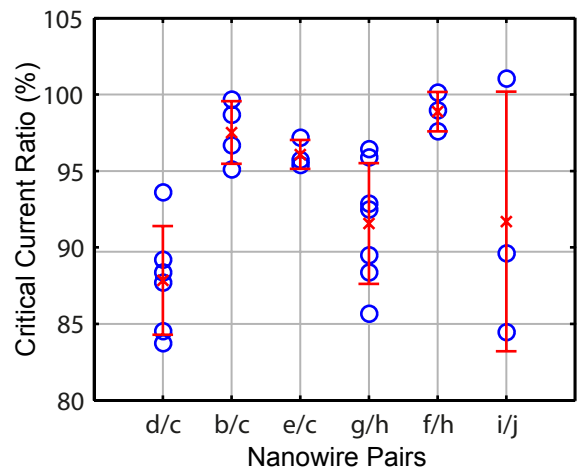} \caption{\label{fig:2}
(Color online) $I_c^\alpha$/$I_c^\beta$ for closely fabricated bent
nanowire pairs labeled by $\alpha$/$\beta$, where $\alpha$ and
$\beta$ correspond to the insets of Fig~\ref{fig:1}. The red error
bars indicate the mean and standard deviation for measurements on
each pair.}
\end{figure}

\begin{figure}[b!]
\centering\includegraphics[width=8.0cm]{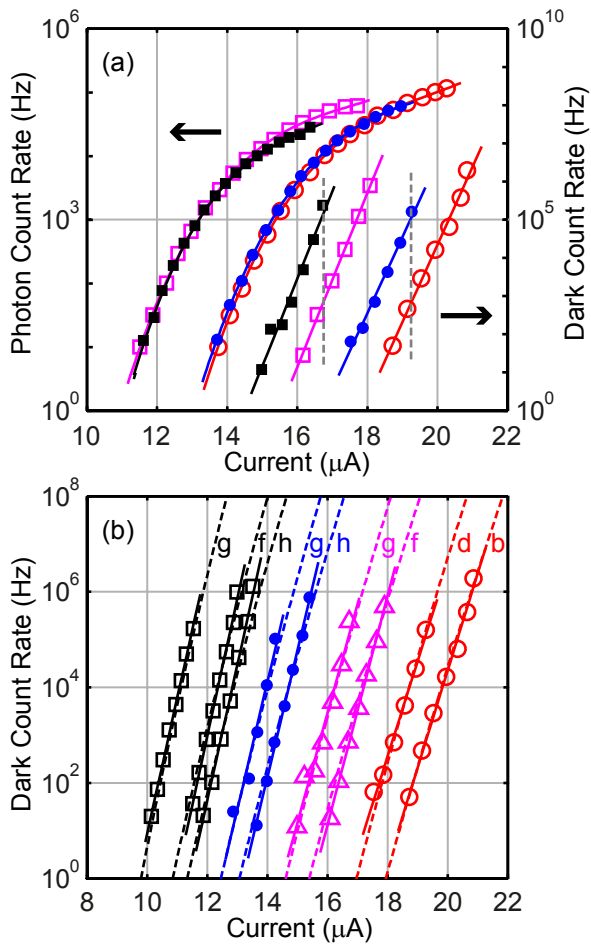} \caption{\label{fig:3}
(Color online) (a) Photo-response and dark count measurements for
samples of pairs d/b and g/f. (b) Dark count measurements for more
samples. Each symbol is for devices on the same chip. The letters
refer to insets of Fig.~\ref{fig:1}. All the lines are for eye
guide.}
\end{figure}

Figure~\ref{fig:2} summarizes the critical current measurements of
the samples. The horizontal axis (i.e. $\alpha/\beta$) specifies a
pair of nanowires using the characters that name the bends in the
insets of Fig.~\ref{fig:1}. The blue circles show the ratio of the
critical currents of the nanowires within the same pair (i.e
$I_c^\alpha$/$I_c^\beta$). The red crosses indicate the mean and the
bars show the standard deviation for measurements of each pair type.
The pairs d/c and g/h show the sharpest bends have considerably
lower critical currents compared to the bigger optimum bends (see
the means). The pairs b/c and f/h show while both of the smallest
and bigger optimum bends work appropriately, bigger bends still
slightly improve the critical current. We attribute this observation
to smaller current density at the inner edge of the bigger designs
and therefore their improved tolerance to fabrication errors. The
pair e/c shows the moderate performance of 45$^\circ$ 90-degree
bends. Finally i/j shows while there is no optimum design for a
180-degree turnaround of 100nm wide strips spaced by 100nm,
circularly rounded bends can improve the reduced critical current by
a considerable factor. We have also confirmed satisfactory operation
of our measurement setup by measuring the critical current of our
devices several times and finding negligible mean normalized error
(equal to $\sim$0.4\%).

The error bars of Fig.~\ref{fig:2} show large variations for
different samples of the same pair. An approximate trend is that the
sharper the bend the larger the variation. We attribute this to the
uncontrollability of the radius of curvature ($\sim$35nm, see yellow
circles of Fig.~\ref{fig:1}) for sharp bends. For the pair i/j the
variation is maximum because $\sim$35nm could almost change device i
to j. This can be a possible explanation for small fabrication yield
of SNSPDs \cite{Natarajan2012Su} where the large number of serially
connected 180-degree turns in a meander makes having at least one
very sharp bend quite possible.

We have also measured the dark counts and photon counts generated by
the nanowires of a given pair. The room temperature end of the fiber
was blocked by a shutter for dark count vs bias current
measurements. Photon counts are measured by exciting the pairs with
weak laser pulses and subtracting the expected dark counts at the
same bias. We ensure single photon sensitivity by checking the
linear proportionally of the photon counts with the number of
incident photons \cite{Akhlaghi2011Non}.

Figure~\ref{fig:3}(a) shows the result for two of our pairs d/b
(circles) and g/f (squares). The sharper bends are shown by filled
symbols. The photo-response of the bends that make a pair is almost
similar for their common range of bias. This is expected as the
devices in a given pair are identical except at the small bending
area. However, at the same bias current, and therefore at the same
quantum efficiency, utilizing an optimum bend can reduce the dark
count rate by orders of magnitude, a significant result for SNSPDs.
We also note that the optimum bends further increase the range of
bias and thus enable operating at higher quantum efficiency (or
longer wavelength).

Illustrated in Fig.~\ref{fig:3}(b) are dark count measurements for
some of the nanowires fabricated on different chips (each symbol is
for devices on the same chip). About $\pm20\%$ variations for
critical current measurements of our optimum bends can be seen.
Investigating the devices under scanning electron microscope, we
have not observed significant dimension changes. Therefore, we
attribute this variations to slight film thickness/quality change
from chip to chip. However, on each chip the trend is the same: the
sharper the bend the smaller the critical current, and the higher
the dark counts.

At the inner edge of a sharp 90-degree turn with radius of curvature
equal to $\sim$35nm, we calculate the density of the sheet current,
$|\textbf{\textit{K}}|$, $\sim$1.7 times higher than the same
density for an optimized bend (smallest possible footprint). So, a
vortex at the edge of a sharp turn faces almost the same barrier as
a vortex at the edge of an optimum bend but at a bias current
$\sim$1.7 times smaller (neglecting radius of curvature effects
\cite{Clem2011Ge} which is reasonable because $\sim$35nm is bigger
than the coherence length). Therefore, assuming vortices overcoming
an edge barrier is the origin of dark counts
\cite{Bulaevskii2011Vo}, we expect having the dark count vs bias
current of a sharp turn to be approximately shifted to smaller
currents by $\sim$1/1.7. However, in none of our nanowires have we
observed such a large shift. The trend of disagreement with this
theory is nevertheless the same as what has been observed for
critical current measurements on wider strips \cite{Henrich2012Ge,
Hortensius2012Cr}.

To conclude, we have explored the possible adverse impact of sharp
turns on SNSPDs through (i) limiting their bias current and thus
limiting their quantum efficiency, and (ii) generating excess dark
counts not generated by straight nanowire segments where photons are
detected. We expect the utilization of optimally designed bends to
further push SNSPDs to more efficient single photon detection at
longer wavelengths while generating less dark counts.


%
%

%

\begin{acknowledgments}
We acknowledge the financial support of OCE, NSERC and IQC. The
authors would like to acknowledge Robin Cantor for helpful comments.
This work was performed in part at the Center for Nanoscale Systems
(CNS), a member of the National Nanotechnology Infrastructure
Network (NNIN), which is supported by the National Science
Foundation under NSF award no. ECS-0335765. CNS is part of Harvard
University.
\end{acknowledgments}


\begin{thebibliography}{14}%
\makeatletter
\providecommand \@ifxundefined [1]{%
 \@ifx{#1\undefined}
}%
\providecommand \@ifnum [1]{%
 \ifnum #1\expandafter \@firstoftwo
 \else \expandafter \@secondoftwo
 \fi
}%
\providecommand \@ifx [1]{%
 \ifx #1\expandafter \@firstoftwo
 \else \expandafter \@secondoftwo
 \fi
}%
\providecommand \natexlab [1]{#1}%
\providecommand \enquote  [1]{``#1''}%
\providecommand \bibnamefont  [1]{#1}%
\providecommand \bibfnamefont [1]{#1}%
\providecommand \citenamefont [1]{#1}%
\providecommand \href@noop [0]{\@secondoftwo}%
\providecommand \href [0]{\begingroup \@sanitize@url \@href}%
\providecommand \@href[1]{\@@startlink{#1}\@@href}%
\providecommand \@@href[1]{\endgroup#1\@@endlink}%
\providecommand \@sanitize@url [0]{\catcode `\\12\catcode
`\$12\catcode
  `\&12\catcode `\#12\catcode `\^12\catcode `\_12\catcode `\%12\relax}%
\providecommand \@@startlink[1]{}%
\providecommand \@@endlink[0]{}%
\providecommand \url  [0]{\begingroup\@sanitize@url \@url }%
\providecommand \@url [1]{\endgroup\@href {#1}{\urlprefix }}%
\providecommand \urlprefix  [0]{URL }%
\providecommand \Eprint [0]{\href }%
\providecommand \doibase [0]{http://dx.doi.org/}%
\providecommand \selectlanguage [0]{\@gobble}%
\providecommand \bibinfo  [0]{\@secondoftwo}%
\providecommand \bibfield  [0]{\@secondoftwo}%
\providecommand \translation [1]{[#1]}%
\providecommand \BibitemOpen [0]{}%
\providecommand \bibitemStop [0]{}%
\providecommand \bibitemNoStop [0]{.\EOS\space}%
\providecommand \EOS [0]{\spacefactor3000\relax}%
\providecommand \BibitemShut  [1]{\csname bibitem#1\endcsname}%
\let\auto@bib@innerbib\@empty
\bibitem [{\citenamefont {Knill}, \citenamefont {Laflamme},\ and\ \citenamefont
  {Milburn}(2001)}]{Knill2001A}%
  \BibitemOpen
  \bibfield  {author} {\bibinfo {author} {\bibfnamefont {E.}~\bibnamefont
  {Knill}}, \bibinfo {author} {\bibfnamefont {R.}~\bibnamefont {Laflamme}}, \
  and\ \bibinfo {author} {\bibfnamefont {G.~J.}\ \bibnamefont {Milburn}},\
  }\href@noop {} {\bibfield  {journal} {\bibinfo  {journal} {Nature}\ }\textbf
  {\bibinfo {volume} {409}},\ \bibinfo {pages} {46--52} (\bibinfo {year}
  {2001})}\BibitemShut {NoStop}%
\bibitem [{\citenamefont {Takesue}\ \emph {et~al.}(2007)\citenamefont
  {Takesue}, \citenamefont {Nam}, \citenamefont {Zhang}, \citenamefont
  {Hadfield}, \citenamefont {Honjo}, \citenamefont {Tamaki},\ and\
  \citenamefont {Yamamoto}}]{Takesue2007Qu}%
  \BibitemOpen
  \bibfield  {author} {\bibinfo {author} {\bibfnamefont {H.}~\bibnamefont
  {Takesue}}, \bibinfo {author} {\bibfnamefont {S.~W.}\ \bibnamefont {Nam}},
  \bibinfo {author} {\bibfnamefont {Q.}~\bibnamefont {Zhang}}, \bibinfo
  {author} {\bibfnamefont {R.~H.}\ \bibnamefont {Hadfield}}, \bibinfo {author}
  {\bibfnamefont {T.}~\bibnamefont {Honjo}}, \bibinfo {author} {\bibfnamefont
  {K.}~\bibnamefont {Tamaki}}, \ and\ \bibinfo {author} {\bibfnamefont
  {Y.}~\bibnamefont {Yamamoto}},\ }\href@noop {} {\bibfield  {journal}
  {\bibinfo  {journal} {Nature Photon.}\ }\textbf {\bibinfo {volume} {1}},\
  \bibinfo {pages} {343--348} (\bibinfo {year} {2007})}\BibitemShut {NoStop}%
\bibitem [{\citenamefont {Grein}\ \emph {et~al.}(2011)\citenamefont {Grein},
  \citenamefont {Kerman}, \citenamefont {Dauler}, \citenamefont {Shatrovoy},
  \citenamefont {Molnar}, \citenamefont {Rosenberg}, \citenamefont {Yoon},
  \citenamefont {Devoe}, \citenamefont {Murphy}, \citenamefont {Robinson},\
  and\ \citenamefont {Boroson}}]{Grein2011De}%
  \BibitemOpen
  \bibfield  {author} {\bibinfo {author} {\bibfnamefont {M.~E.}\ \bibnamefont
  {Grein}}, \bibinfo {author} {\bibfnamefont {A.~J.}\ \bibnamefont {Kerman}},
  \bibinfo {author} {\bibfnamefont {E.~A.}\ \bibnamefont {Dauler}}, \bibinfo
  {author} {\bibfnamefont {O.}~\bibnamefont {Shatrovoy}}, \bibinfo {author}
  {\bibfnamefont {R.~J.}\ \bibnamefont {Molnar}}, \bibinfo {author}
  {\bibfnamefont {D.}~\bibnamefont {Rosenberg}}, \bibinfo {author}
  {\bibfnamefont {J.}~\bibnamefont {Yoon}}, \bibinfo {author} {\bibfnamefont
  {C.~E.}\ \bibnamefont {Devoe}}, \bibinfo {author} {\bibfnamefont {D.~V.}\
  \bibnamefont {Murphy}}, \bibinfo {author} {\bibfnamefont {B.~S.}\
  \bibnamefont {Robinson}}, \ and\ \bibinfo {author} {\bibfnamefont {D.~M.}\
  \bibnamefont {Boroson}},\ }\href@noop {} {\bibfield  {journal} {\bibinfo
  {journal} {2011 International Conference on Space Optical Systems and
  Applications, ICSOS'11}\ ,\ \bibinfo {pages} {78--82}} (\bibinfo {year}
  {2011})}\BibitemShut {NoStop}%
\bibitem [{\citenamefont {Zhang}\ \emph {et~al.}(2003)\citenamefont {Zhang},
  \citenamefont {Boiadjieva}, \citenamefont {Chulkova}, \citenamefont
  {Deslandes}, \citenamefont {Gol'tsman}, \citenamefont {Korneev},
  \citenamefont {Kouminov}, \citenamefont {Leibowitz}, \citenamefont {Lo},
  \citenamefont {Malinsky}, \citenamefont {Okunev}, \citenamefont {Pearlman},
  \citenamefont {Slysz}, \citenamefont {Smirnov}, \citenamefont {Tsao},
  \citenamefont {Verevkin}, \citenamefont {Voronov}, \citenamefont {Wilsher},\
  and\ \citenamefont {Sobolewski}}]{Zhang2003No}%
  \BibitemOpen
  \bibfield  {author} {\bibinfo {author} {\bibfnamefont {J.}~\bibnamefont
  {Zhang}}, \bibinfo {author} {\bibfnamefont {N.}~\bibnamefont {Boiadjieva}},
  \bibinfo {author} {\bibfnamefont {G.}~\bibnamefont {Chulkova}}, \bibinfo
  {author} {\bibfnamefont {H.}~\bibnamefont {Deslandes}}, \bibinfo {author}
  {\bibfnamefont {G.~N.}\ \bibnamefont {Gol'tsman}}, \bibinfo {author}
  {\bibfnamefont {A.}~\bibnamefont {Korneev}}, \bibinfo {author} {\bibfnamefont
  {P.}~\bibnamefont {Kouminov}}, \bibinfo {author} {\bibfnamefont
  {M.}~\bibnamefont {Leibowitz}}, \bibinfo {author} {\bibfnamefont
  {W.}~\bibnamefont {Lo}}, \bibinfo {author} {\bibfnamefont {R.}~\bibnamefont
  {Malinsky}}, \bibinfo {author} {\bibfnamefont {O.}~\bibnamefont {Okunev}},
  \bibinfo {author} {\bibfnamefont {A.}~\bibnamefont {Pearlman}}, \bibinfo
  {author} {\bibfnamefont {W.}~\bibnamefont {Slysz}}, \bibinfo {author}
  {\bibfnamefont {K.}~\bibnamefont {Smirnov}}, \bibinfo {author} {\bibfnamefont
  {C.}~\bibnamefont {Tsao}}, \bibinfo {author} {\bibfnamefont {A.}~\bibnamefont
  {Verevkin}}, \bibinfo {author} {\bibfnamefont {B.}~\bibnamefont {Voronov}},
  \bibinfo {author} {\bibfnamefont {K.}~\bibnamefont {Wilsher}}, \ and\
  \bibinfo {author} {\bibfnamefont {R.}~\bibnamefont {Sobolewski}},\
  }\href@noop {} {\bibfield  {journal} {\bibinfo  {journal} {Electron. Lett.}\
  }\textbf {\bibinfo {volume} {39}},\ \bibinfo {pages} {1086--1088} (\bibinfo
  {year} {2003})}\BibitemShut {NoStop}%
\bibitem [{\citenamefont {Stevens}\ \emph {et~al.}(2006)\citenamefont
  {Stevens}, \citenamefont {Hadfield}, \citenamefont {Schwall}, \citenamefont
  {Nam}, \citenamefont {Mirin},\ and\ \citenamefont {Gupta}}]{Stevens2006Fa}%
  \BibitemOpen
  \bibfield  {author} {\bibinfo {author} {\bibfnamefont {M.~J.}\ \bibnamefont
  {Stevens}}, \bibinfo {author} {\bibfnamefont {R.~H.}\ \bibnamefont
  {Hadfield}}, \bibinfo {author} {\bibfnamefont {R.~E.}\ \bibnamefont
  {Schwall}}, \bibinfo {author} {\bibfnamefont {S.~W.}\ \bibnamefont {Nam}},
  \bibinfo {author} {\bibfnamefont {R.~P.}\ \bibnamefont {Mirin}}, \ and\
  \bibinfo {author} {\bibfnamefont {J.~A.}\ \bibnamefont {Gupta}},\ }\href@noop
  {} {\bibfield  {journal} {\bibinfo  {journal} {Appl. Phys. Lett.}\ }\textbf
  {\bibinfo {volume} {89}} (\bibinfo {year} {2006})}\BibitemShut {NoStop}%
\bibitem [{\citenamefont {Hadfield}(2009)}]{Hadfield2009S}%
  \BibitemOpen
  \bibfield  {author} {\bibinfo {author} {\bibfnamefont {R.~H.}\ \bibnamefont
  {Hadfield}},\ }\href@noop {} {\bibfield  {journal} {\bibinfo  {journal}
  {Nature Photonics}\ }\textbf {\bibinfo {volume} {3}},\ \bibinfo {pages}
  {696--705} (\bibinfo {year} {2009})}\BibitemShut {NoStop}%
\bibitem [{\citenamefont {Akhlaghi}\ and\ \citenamefont
  {Majedi}(2012)}]{Akhlaghi2012Ga}%
  \BibitemOpen
  \bibfield  {author} {\bibinfo {author} {\bibfnamefont {M.~K.}\ \bibnamefont
  {Akhlaghi}}\ and\ \bibinfo {author} {\bibfnamefont {A.~H.}\ \bibnamefont
  {Majedi}},\ }\href@noop {} {\bibfield  {journal} {\bibinfo  {journal} {Opt.
  Express}\ }\textbf {\bibinfo {volume} {20}},\ \bibinfo {pages} {1608--1616}
  (\bibinfo {year} {2012})}\BibitemShut {NoStop}%
\bibitem [{\citenamefont {Natarajan}, \citenamefont {Tanner},\ and\
  \citenamefont {Hadfield}(2012)}]{Natarajan2012Su}%
  \BibitemOpen
  \bibfield  {author} {\bibinfo {author} {\bibfnamefont {C.~M.}\ \bibnamefont
  {Natarajan}}, \bibinfo {author} {\bibfnamefont {M.~G.}\ \bibnamefont
  {Tanner}}, \ and\ \bibinfo {author} {\bibfnamefont {R.~H.}\ \bibnamefont
  {Hadfield}},\ }\href@noop {} {\bibfield  {journal} {\bibinfo  {journal}
  {Superconductor Science and Technology}\ }\textbf {\bibinfo {volume} {25}},\
  \bibinfo {pages} {063001} (\bibinfo {year} {2012})}\BibitemShut {NoStop}%
\bibitem [{\citenamefont {Clem}\ and\ \citenamefont
  {Berggren}(2011)}]{Clem2011Ge}%
  \BibitemOpen
  \bibfield  {author} {\bibinfo {author} {\bibfnamefont {J.~R.}\ \bibnamefont
  {Clem}}\ and\ \bibinfo {author} {\bibfnamefont {K.~K.}\ \bibnamefont
  {Berggren}},\ }\href@noop {} {\bibfield  {journal} {\bibinfo  {journal}
  {Phys. Rev. B}\ }\textbf {\bibinfo {volume} {84}},\ \bibinfo {pages} {174510}
  (\bibinfo {year} {2011})}\BibitemShut {NoStop}%
\bibitem [{\citenamefont {Hortensius}\ \emph {et~al.}(2012)\citenamefont
  {Hortensius}, \citenamefont {Driessen}, \citenamefont {Klapwijk},
  \citenamefont {Berggren},\ and\ \citenamefont {Clem}}]{Hortensius2012Cr}%
  \BibitemOpen
  \bibfield  {author} {\bibinfo {author} {\bibfnamefont {H.~L.}\ \bibnamefont
  {Hortensius}}, \bibinfo {author} {\bibfnamefont {E.~F.~C.}\ \bibnamefont
  {Driessen}}, \bibinfo {author} {\bibfnamefont {T.~M.}\ \bibnamefont
  {Klapwijk}}, \bibinfo {author} {\bibfnamefont {K.~K.}\ \bibnamefont
  {Berggren}}, \ and\ \bibinfo {author} {\bibfnamefont {J.~R.}\ \bibnamefont
  {Clem}},\ }\href@noop {} {\bibfield  {journal} {\bibinfo  {journal} {Applied
  Physics Letters}\ }\textbf {\bibinfo {volume} {100}},\ \bibinfo {pages}
  {182602} (\bibinfo {year} {2012})}\BibitemShut {NoStop}%
\bibitem [{\citenamefont {Henrich}\ \emph {et~al.}()\citenamefont {Henrich},
  \citenamefont {Reichensperger}, \citenamefont {Hofherr}, \citenamefont
  {Ilin}, \citenamefont {Siegel}, \citenamefont {Semenov}, \citenamefont
  {Zotova},\ and\ \citenamefont {Vodolazov}}]{Henrich2012Ge}%
  \BibitemOpen
  \bibfield  {author} {\bibinfo {author} {\bibfnamefont {D.}~\bibnamefont
  {Henrich}}, \bibinfo {author} {\bibfnamefont {P.}~\bibnamefont
  {Reichensperger}}, \bibinfo {author} {\bibfnamefont {M.}~\bibnamefont
  {Hofherr}}, \bibinfo {author} {\bibfnamefont {K.}~\bibnamefont {Ilin}},
  \bibinfo {author} {\bibfnamefont {M.}~\bibnamefont {Siegel}}, \bibinfo
  {author} {\bibfnamefont {A.}~\bibnamefont {Semenov}}, \bibinfo {author}
  {\bibfnamefont {A.}~\bibnamefont {Zotova}}, \ and\ \bibinfo {author}
  {\bibfnamefont {D.~Y.}\ \bibnamefont {Vodolazov}},\ }\href@noop {} {\bibinfo
  {journal} {arXiv:1204.0616v1 (unpublished)}\ }\BibitemShut {NoStop}%
\bibitem [{\citenamefont {Kerman}\ \emph {et~al.}(2009)\citenamefont {Kerman},
  \citenamefont {Yang}, \citenamefont {Molnar}, \citenamefont {Dauler},\ and\
  \citenamefont {Berggren}}]{Kerman2009El}%
  \BibitemOpen
\bibfield  {journal} {  }\bibfield  {author} {\bibinfo {author} {\bibfnamefont
  {A.~J.}\ \bibnamefont {Kerman}}, \bibinfo {author} {\bibfnamefont {J.~K.~W.}\
  \bibnamefont {Yang}}, \bibinfo {author} {\bibfnamefont {R.~J.}\ \bibnamefont
  {Molnar}}, \bibinfo {author} {\bibfnamefont {E.~A.}\ \bibnamefont {Dauler}},
  \ and\ \bibinfo {author} {\bibfnamefont {K.~K.}\ \bibnamefont {Berggren}},\
  }\href@noop {} {\bibfield  {journal} {\bibinfo  {journal} {Phys. Rev. B}\
  }\textbf {\bibinfo {volume} {79}} (\bibinfo {year} {2009})}\BibitemShut
  {NoStop}%
\bibitem [{\citenamefont {Akhlaghi}, \citenamefont {Majedi},\ and\
  \citenamefont {Lundeen}(2011)}]{Akhlaghi2011Non}%
  \BibitemOpen
  \bibfield  {author} {\bibinfo {author} {\bibfnamefont {M.~K.}\ \bibnamefont
  {Akhlaghi}}, \bibinfo {author} {\bibfnamefont {A.~H.}\ \bibnamefont
  {Majedi}}, \ and\ \bibinfo {author} {\bibfnamefont {J.~S.}\ \bibnamefont
  {Lundeen}},\ }\href@noop {} {\bibfield  {journal} {\bibinfo  {journal} {Opt.
  Express}\ }\textbf {\bibinfo {volume} {19}},\ \bibinfo {pages} {21305--21312}
  (\bibinfo {year} {2011})}\BibitemShut {NoStop}%
\bibitem [{\citenamefont {Bulaevskii}\ \emph {et~al.}(2011)\citenamefont
  {Bulaevskii}, \citenamefont {Graf}, \citenamefont {Batista},\ and\
  \citenamefont {Kogan}}]{Bulaevskii2011Vo}%
  \BibitemOpen
  \bibfield  {author} {\bibinfo {author} {\bibfnamefont {L.~N.}\ \bibnamefont
  {Bulaevskii}}, \bibinfo {author} {\bibfnamefont {M.~J.}\ \bibnamefont
  {Graf}}, \bibinfo {author} {\bibfnamefont {C.~D.}\ \bibnamefont {Batista}}, \
  and\ \bibinfo {author} {\bibfnamefont {V.~G.}\ \bibnamefont {Kogan}},\
  }\href@noop {} {\bibfield  {journal} {\bibinfo  {journal} {Phys. Rev. B}\
  }\textbf {\bibinfo {volume} {83}},\ \bibinfo {pages} {144526} (\bibinfo
  {year} {2011})}\BibitemShut {NoStop}%
\end{thebibliography}

%

\end{document}